\documentclass[aps,physrev,reprint,superscriptaddress]{revtex4-2}
\usepackage{graphicx}
\usepackage{hyperref}
\hypersetup{colorlinks=true, linkcolor=blue, anchorcolor=blue, citecolor=blue, urlcolor=blue}
\usepackage{multirow}
\usepackage{amsmath}
\usepackage{xcolor}
\usepackage{amssymb}

\begin{document}

%\preprint{}

\title{Exceptionally high carrier mobility in hexagonal diamond}


\author{Zirui He}
%\thanks{}
\affiliation{College of Smart Materials and Future Energy, Fudan University, Shanghai 200433, China}
\affiliation{Shanghai Advanced Silicon Technology Co., Ltd., Shanghai 201616, China}

\author{Shang-Peng Gao}
\email{gaosp@fudan.edu.cn}
\affiliation{College of Smart Materials and Future Energy, Fudan University, Shanghai 200433, China}
\affiliation{Shanghai Advanced Silicon Technology Co., Ltd., Shanghai 201616, China}

\author{Meng Chen}
\email{mchen@ast.com.cn}
\affiliation{Shanghai Advanced Silicon Technology Co., Ltd., Shanghai 201616, China}


%\date{\today}

\begin{abstract}
Hexagonal diamond (h-diamond), or Lonsdaleite, is a promising wide-bandgap semiconductor known for its high thermal conductivity and hardness.
Based on \textit{ab initio} calculations, we demonstrate its exceptionally high carrier mobilities.
At room temperature, the hole mobilities along the $\perp c$ and $\parallel c$ directions are 6000 and 6024~cm$^{2}$V$^{-1}$s$^{-1}$, respectively, while the corresponding electron mobilities reach 12339 and 28473~cm$^{2}$V$^{-1}$s$^{-1}$.
These values are significantly superior to those of most known semiconductors, including cubic diamond.
The small effective masses in h-diamond are comparable to those in the cubic phase, which cannot explain its substantially higher mobilities.
Instead, two underlying mechanisms are uncovered. 
First, selection rules enforced by the symmetry of h-diamond significantly suppress scattering, particularly for transverse acoustic phonons, which predominate in the cubic phase around room temperature.
Secondly, the spatial mismatch between the electronic wavefunctions and phonon-induced scattering potentials leads to real-space electron-phonon decoupling, which manifests as the suppression of out-of-plane polarised longitudinal acoustic scattering for holes, and a systematic weakening of acoustic scattering for electrons.
\end{abstract}

%\keywords{}

\maketitle

\section{Introduction}
Diamond, long prized as a gemstone, is also regarded as the ``ultimate'' wide-bandgap semiconductor material because of its excellent physical properties~\cite{Wort2008}.
While most semiconductors may show good performance in certain properties, they often face inherent limitations in other critical characteristics.
For example, GaAs has high electron mobility yet low hole mobility and thermal conductivity~\cite{Blakemore1982}; $\beta$-Ga$_2$O$_3$ has an ultrawide bandgap and large breakdown field, whereas its carrier mobility and thermal conductivity are unsatisfactory~\cite{Galazka2018}.
In contrast, diamond offers a unique combination of a series of key properties, including its  high carrier mobility, high thermal conductivity, ultrawide bandgap, and high breakdown field~\cite{Wort2008}.
These intrinsic characteristics position it at the frontier of future electronics operating under the most extreme conditions of high power, frequency, and temperature~\cite{Shikata2016}.

In addition to the well-known cubic structure, diamond can also crystallise in the hexagonal phase.
Hexagonal diamond (h-diamond), also referred to as Lonsdaleite~\cite{Frondel1967}, was both discovered in meteorites~\cite{Hanneman1967} and experimentally synthesised~\cite{Bundy1967} in the 1960s.
In contrast to cubic diamond (c-diamond) with a stacking sequence of ABCABC{\dots} along the $\langle111\rangle$ directions, it exhibits an ABAB{\dots} stacking.
Although far less explored than the well-known cubic phase, existing studies have shown that the properties of h-diamond are comparable to, or even superior to, those of c-diamond.
For example, Pan \textit{et al.}~\cite{Pan2009} revealed a 58\% higher indentation strength of h-diamond than that of its cubic counterpart.
Regarding lattice thermal conductivity, studies on h-diamond report scattered values~\cite{Zhang2025,Shi2021,Chakraborty2018}, which are approximately 20--30\% lower than that of c-diamond, yet still considerably higher than those of most known semiconductors.

In this work, motivated by the recent success in the experimental synthesis of its highly-ordered bulk samples~\cite{Yang2025,Lai2026}, we thoroughly investigate the carrier mobility of h-diamond using fully \textit{ab initio} calculations, revealing that its mobility significantly exceeds that of most known semiconductor materials.
Based on group theory and the spatial distributions of charge density and scattering potential, we further uncover two crucial mechanisms as the physical origins of its high mobility, thereby guiding the discovery and design of novel high-mobility materials.

\section{Results}

\subsection{Phonon-limited carrier mobility}

\begin{figure*}
    \centering
    \includegraphics[width=\linewidth]{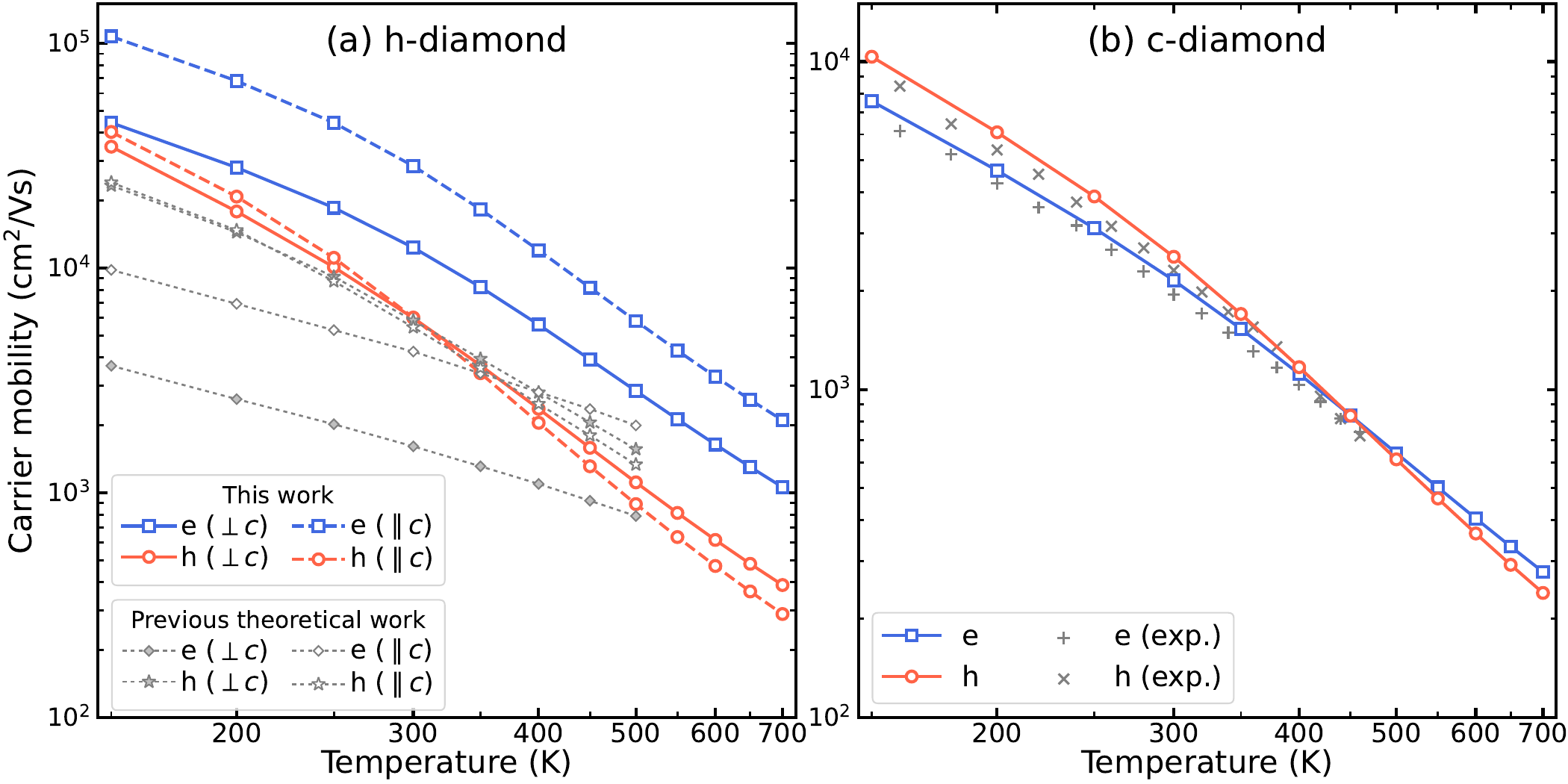}
    \caption{Carrier mobilities of (a) h-diamond and (b) c-diamond, as functions of temperature. Previous theoretical data in (a) and experimental data in (b) are taken from Refs.~\onlinecite{Zheng2024} and \onlinecite{Gabrysch2011}, respectively.}
    \label{fig:mob-temp}
\end{figure*}

The calculated carrier mobilities of both phases are shown in Fig.~\ref{fig:mob-temp}.
For c-diamond, the room-temperature hole mobility ($\mu_\mathrm{h}$) and electron mobility ($\mu_\mathrm{e}$) are predicted to be 2540 and 2152~cm$^2$V$^{-1}$s$^{-1}$, respectively, in good agreement with experimental values~\cite{Gabrysch2011} and other theoretical data ($\mu_\mathrm{h}$ = 2500~cm$^2$V$^{-1}$s$^{-1}$~\cite{Macheda2018}).
Since the $G_0W_0$ quasi-particle levels were used in our work (see the Methods section), here we also performed complementary calculations with the standard generalised gradient approximation (GGA) Kohn-Sham band structure, to enable a better comparison with previous computational work such as Ref.~\onlinecite{Ponce2021}.
We obtained hole and electron mobilities of 2151 and 1680~cm$^2$V$^{-1}$s$^{-1}$, respectively, in good agreement with those reported in Ref.~\onlinecite{Ponce2021} ($\mu_\mathrm h$ = 2290~cm$^2$V$^{-1}$s$^{-1}$ and $\mu_\mathrm e$ = 1664~cm$^2$V$^{-1}$s$^{-1}$), thus further confirming the robustness of our computational method and the reliability of the numerical convergence.

Regarding h-diamond, its carrier mobilities are found to be much higher.
At 300~K, the hole mobilities are $\mu_\mathrm{h}^{\perp c}$ = 6000 and $\mu_\mathrm{h}^{\parallel c}$ = 6024~cm$^2$V$^{-1}$s$^{-1}$.
The electron mobilities are even higher, with $\mu_\mathrm{e}^{\perp c}$ = 12339 and $\mu_\mathrm{e}^{\parallel c}$ = 28473~cm$^2$V$^{-1}$s$^{-1}$.
These values are superior to the typical mobility of most known high-mobility semiconductors, for example, Ge for holes ($\mu_\mathrm{h}$ = 1820~cm$^2$V$^{-1}$s$^{-1}$~\cite{Morin1954}) and GaAs for electrons ($\mu_\mathrm{e}$ = 8900~cm$^2$V$^{-1}$s$^{-1}$~\cite{Hicks1969}), also overcoming the limitation of most conventional semiconductors where only one carrier type exhibits high mobility.
While the predicted hole mobility agrees well with the result in Ref.~\onlinecite{Zheng2024}, our calculated electron mobility significantly exceeds the reported value.
We carried out cross-verification using another interpolation scheme, which confirmed the exceptional charge transport properties of h-diamond [see Table~S1 in the Supplementary Information (SI)].
Since h-diamond has also been predicted to have a room-temperature thermal conductivity of nearly 2000~Wm$^{-1}$K$^{-1}$~\cite{Chakraborty2018,Zhang2025}, it may outperform other novel semiconductors such as BAs, which has attracted considerable attention in recent years due to its high ambipolar mobility (approximately 1600~cm$^2$V$^{-1}$s$^{-1}$)~\cite{Yue2022,Shin2022} and high thermal conductivity (approximately 1300~Wm$^{-1}$K$^{-1}$)~\cite{Kang2018,Li2018,Tian2018}.
Additionally, our $G_0W_0$ calculation yields a wide bandgap of 4.55~eV for h-diamond, and a molecular dynamics (MD) simulation at 1000~K indicates that the system is thermally and structurally stable at elevated temperatures (see Fig.~S1).
These combined properties suggest h-diamond as a promising candidate for applications in high-power and high-frequency electronics.

To pave the way for the high-temperature application of h-diamond, we examined the temperature dependence of its carrier mobility.
Phonon-limited carrier mobility typically follows a power-law relationship with temperature, \textit{i.e.}, $\mu\propto T^{-\alpha}$, if the scattering mechanism remains unchanged.
However, as shown in Fig.~\ref{fig:mob-temp}(a), the temperature dependence of carrier mobility in h-diamond cannot be well described by a single power-law exponent $\alpha$ over the range of 150--700~K, suggesting a transition in the predominant scattering mechanism, which will be analysed in the subsequent section.

\subsection{Predominant electron-phonon scattering mechanism}

\begin{figure*}
    \centering
    \includegraphics[width=\linewidth]{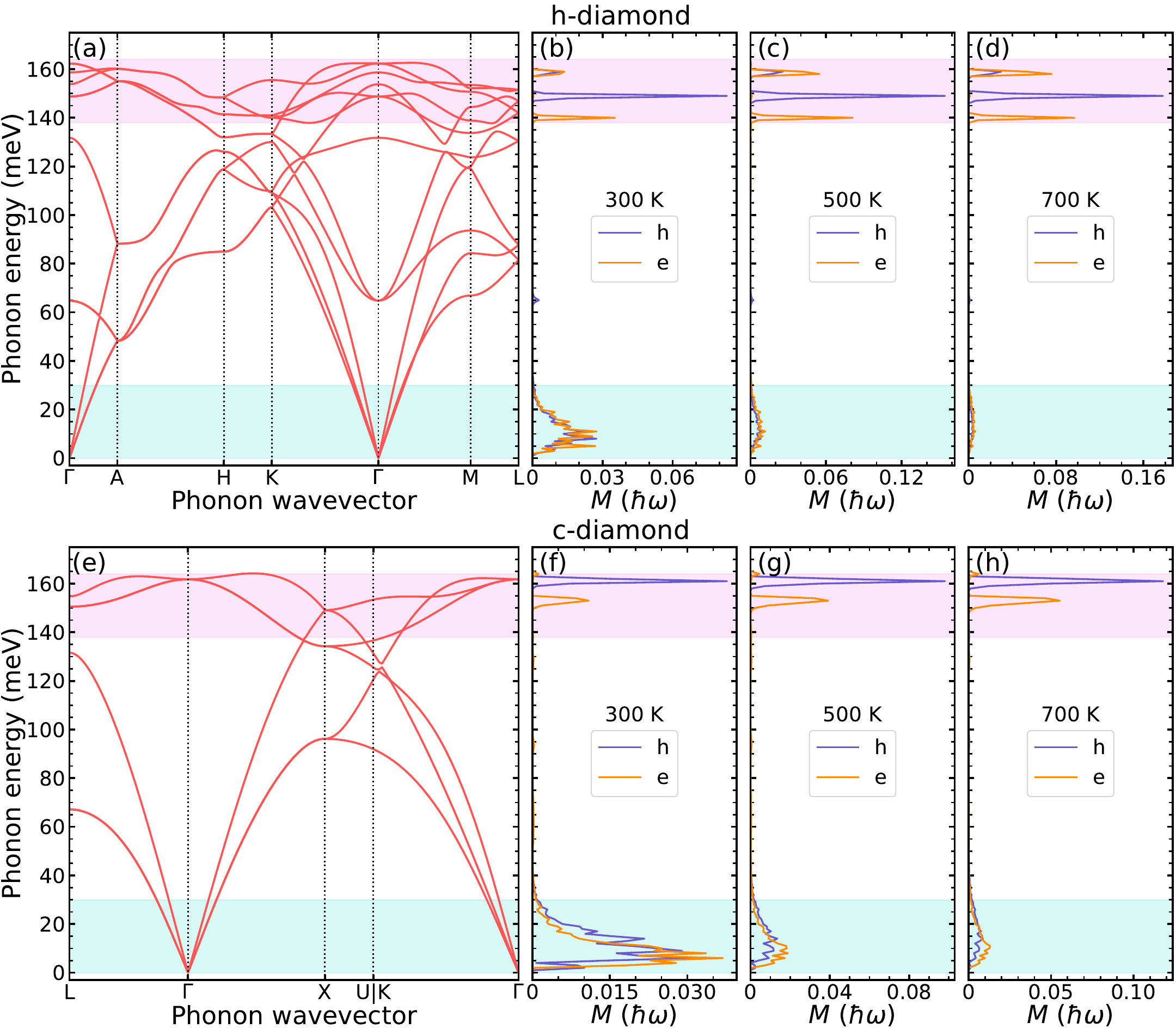}
    \caption{Calculated phonon dispersion and mobility reduction index $M(\hbar\omega)$ at various temperatures for holes and electrons. A larger value of $M (\hbar\omega)$ indicates a greater impact of phonons with energy $\hbar\omega$ on limiting the carrier mobility. For h-diamond, the values along three Cartesian directions are averaged. Two distinct frequency ranges corresponding to significant contributions to carrier scattering by acoustic phonons (marked in turquoise) and optical phonons (marked in violet) are highlighted.} 
    \label{fig:mob-freq}
\end{figure*}

\begin{figure*}
    \centering
    \includegraphics[width=0.83\linewidth]{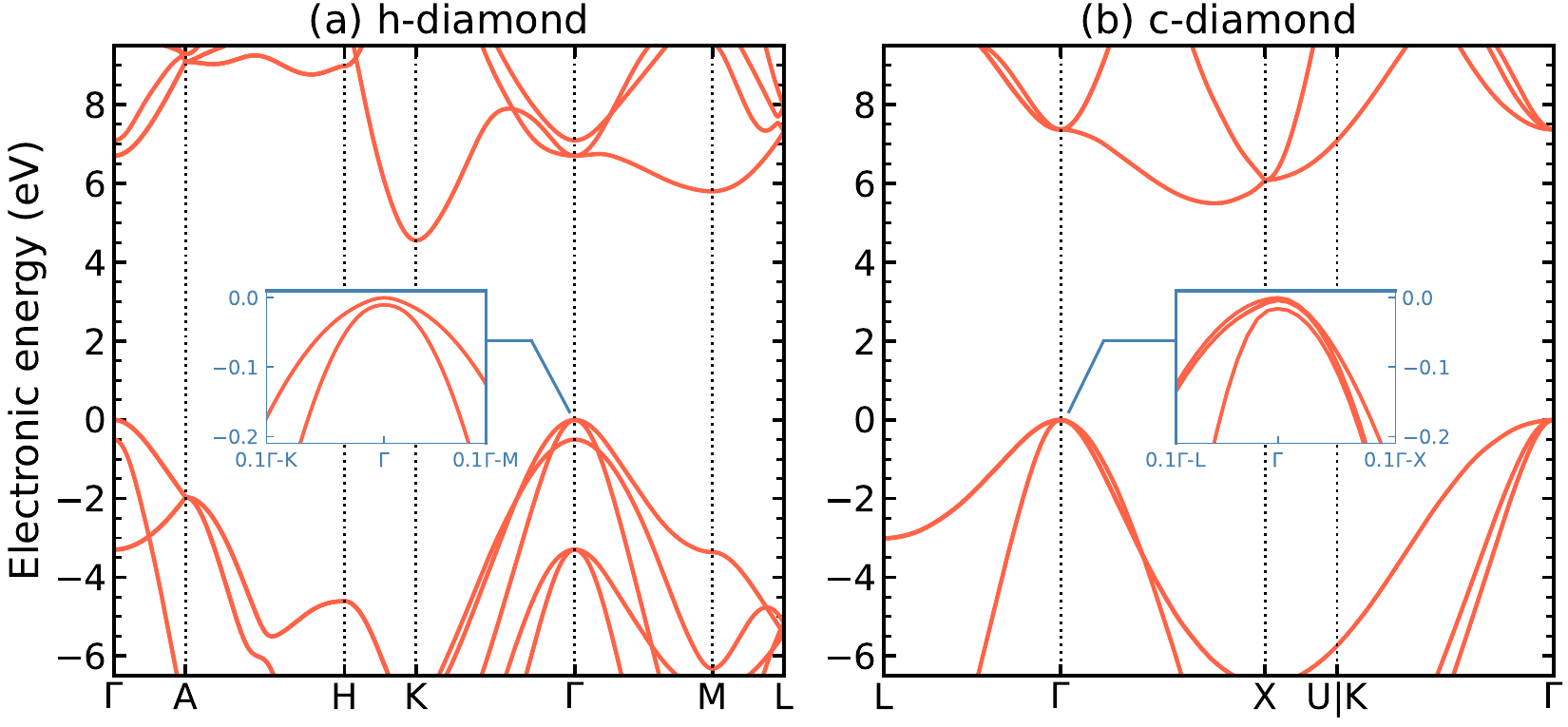}
    \caption{$G_0W_0$ quasi-particle band structures of (a) h-diamond and (b) c-diamond. The fundamental gaps are (a) 4.55~eV and (b) 5.50~eV. Zoom-in views near the VBM are shown in the insets to better resolve the small spin-orbit coupling (SOC)-induced splitting.}
    \label{fig:band}
\end{figure*}

Carrier mobility is governed by scattering mechanisms, an in-depth analysis of which can provide a better understanding of carrier transport behaviour at the microscopic level.
In terms of electron-phonon (e-ph) interaction, since h-diamond is a non-polar semiconductor, deformation potential scattering is expected to play a central role, which can be induced by either acoustic or optical phonons.
We note that the analysis of mode-specific contributions to mobility reduction can be achieved through a spectral decomposition of inverse mobility under a generalised Matthiessen's rule framework, as successfully demonstrated in Ref.~\onlinecite{Ponce2020}.
In this work, we offer an alternative perspective by employing a numerical filtering approach.
We define the ``mobility reduction index'' $M$ as a function of phonon energy $\hbar\omega$ [Eq.~(S6)].
A larger value of $M(\hbar\omega)$ indicates a greater reduction in carrier mobility due to phonons with an energy around $\hbar\omega$.
As shown in Fig.~\ref{fig:mob-freq}, for both phases, carrier scattering is primarily induced by phonons in two energy ranges.
The lower range is below 30~meV, attributed to acoustic deformation potential (ADP) scattering.
The higher range is between 138 and 160~meV, corresponding to optical modes.
At relatively low temperatures (around 300~K), ADP is the predominant scattering mechanism.
Nevertheless, the relative contribution of optical deformation potential (ODP) scattering increases significantly at higher temperatures, explaining the observed deviation from the ideal power-law relationship in Fig.~\ref{fig:mob-temp}.
This finding is fully consistent with Ref.~\onlinecite{Macheda2018}, which studied hole transport in c-diamond and demonstrated that the contribution of the zone-centre optical modes is minor at 300~K, but becomes comparable to that of acoustic phonons at around 500~K.

\subsection{Effect of band structure on carrier transport}
Given the same elemental composition and similar $sp^3$ bonding character of both phases, it is scientifically intriguing to elucidate the microscopic mechanism that gives rise to the substantially higher mobility in h-diamond.
A natural consideration is the difference in band structure.
To rigorously test this hypothesis, we systematically analyse several potential band-related factors, including effective mass, density of states (DOS), and interband or intervalley scattering channels.
The results reveal that the band structure plays an almost identical role in both phases, thereby ruling it out as the source of the mobility difference.
The detailed analyses are provided below.

First, the effective mass is usually regarded as a key factor.
A smaller effective mass typically indicates higher band velocities and a lower density of states, which lead to faster carrier transport and reduced scattering phase space, respectively, thereby increasing mobility.
In order to enable a direct comparison of effective mass between the hexagonal and cubic lattices, we calculated the ``mobility effective masses''~\cite{Ponce2021} to quantify the effect of band curvature on carrier mobility, as listed in Table~\ref{tab:mass}.
We also evaluated the DOS (per valley) near band edges, as shown in Fig.~S2.
Both quantities of h-diamond are comparable to those of c-diamond, indicating that the small effective mass is an important contributor to  the high mobility in h-diamond, considering that the high hole mobility in c-diamond is largely attributed to its small effective mass~\cite{Yang2024}.
However, it cannot further explain the fact that the mobility in h-diamond is significantly higher than that in c-diamond.

\begin{table}
\caption{\label{tab:mass}Calculated mobility effective masses, evaluated at 300~K and given in units of the electron rest mass. The definition is given in Eq.~(S8).} 
\begin{ruledtabular}
\begin{tabular}{lcccc}
 \multirow{2}{*}{Carrier} & \multicolumn{3}{c}{h-diamond} & \multirow{2}{*}{c-diamond} \\
 
 &  $\perp c$ & $\parallel c$ & average &   \\
 \hline
 Hole & 0.477 & 0.678 & 0.530 & 0.489 \\
 Electron & 0.462 & 0.261 & 0.368 & 0.371 \\
\end{tabular}
\end{ruledtabular}
\end{table}

\begin{table}
\caption{\label{tab:ii-mob}Calculated impurity-limited carrier mobility at 300~K (in cm$^2$V$^{-1}$s$^{-1}$). The ionised impurities are singly charged, with a concentration of $10^{17}$~cm$^{-3}$.} 
\begin{ruledtabular}
\begin{tabular}{lccc}
 \multirow{2}{*}{Carrier} & \multicolumn{2}{c}{h-diamond} & \multirow{2}{*}{c-diamond} \\
 
 &  $\perp c$ & $\parallel c$ &   \\
 \hline
 Hole & 1904 & 1521  & 2121 \\
 Electron & 2005 & 2421 & 1994 \\
\end{tabular}
\end{ruledtabular}
\end{table}

Another possible difference is the scattering processes that involve multiple bands or valleys.
As shown in Fig.~\ref{fig:band}, the heavy-hole (hh) and light-hole (lh) bands (referred to as the highest and second highest valence bands, respectively) are close in energy, leading to interband scattering of holes.
Regarding the conduction bands, the conduction band minimum (CBM) is away from the Brillouin zone (BZ) centre, resulting in intervalley scattering between equivalent valleys.

For holes, the crystal-field splitting in h-diamond opens an appreciable energy gap between the valence band maximum (VBM) and the split-off hole (sh) band, whereas in c-diamond, this gap is only 16~meV at the $\Gamma$ point.
It is thus natural to consider whether the lower hole mobility in c-diamond arises from the additional interband scattering channels that involve the sh band.
However, the DOS near the VBM is actually comparable in both phases, as shown in Fig.~S2.
To provide a quantitative analysis, we computed the room-temperature hole mobility with the sh band artificially excluded from the calculation.
The result shows a slight increase in hole mobility of approximately 2\%.

\begin{figure*}
    \centering
    \includegraphics[width=\linewidth]{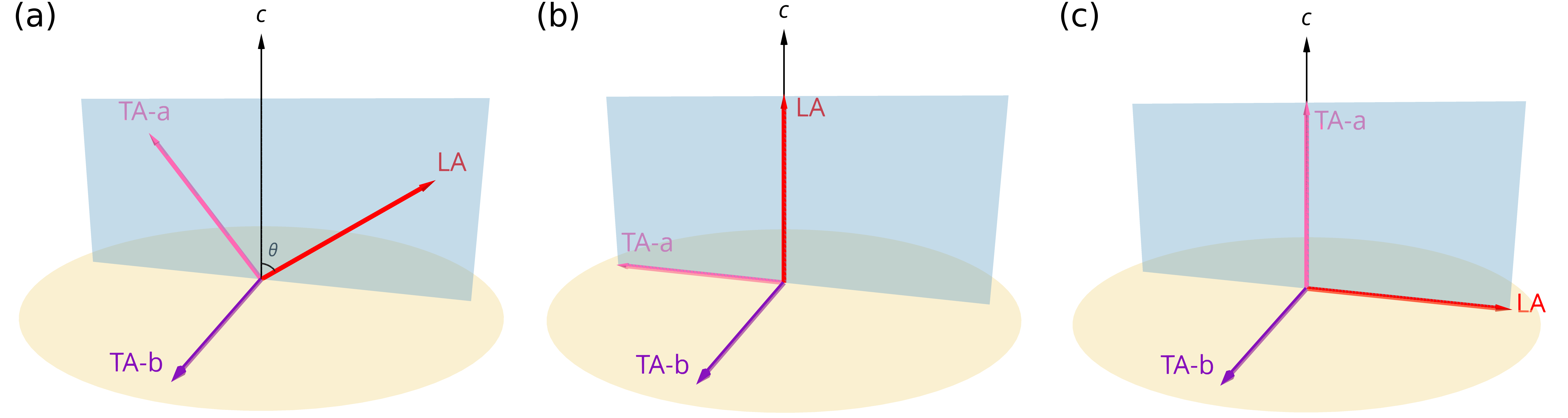}
    \caption{Schematic definition of the TA-a and TA-b phonon modes in h-diamond. Coloured vectors denote the polarisation vectors, with the phonon wavevector $\mathbf q$ parallel to the LA polarisation vector (red). The yellow plane represents the $k_z=0$ plane. The blue plane contains both the $c$-axis and $\mathbf{q}$, with $\theta$ as the angle between them. The polarisation vectors of TA-a and TA-b always lie within the blue and yellow planes, respectively. (a) General case, (b) degenerate in-plane modes for $\mathbf q \parallel c$, and (c) out-of-plane TA-a mode for $\mathbf q \perp c$.}
    \label{fig:hex-TA}
\end{figure*}

Regarding electrons, the CBM of c-diamond lies at the $\Delta$ points and has a larger multiplicity than the CBM of h-diamond (located at $K$). 
To compare the relative contributions of intervalley scattering in both materials, we calculated the room-temperature electron mobility that involves only scattering induced by phonons with wavevector lengths below a specific cutoff, \textit{i.e.}, $|\mathbf q |_\mathrm{cut}$.
Since intravalley scattering is mediated by phonons with small $|\mathbf q|$, and the CBM valleys are well separated from their equivalent counterparts in both phases, the mobility-$|\mathbf q|_\mathrm{cut}$ relationship directly decomposes the contributions of intervalley and intravalley scattering.
As shown in Fig.~S3, intravalley scattering significantly dominates in h-diamond.
In c-diamond, despite the multiple intervalley scattering paths (including $g$-processes and $f$-processes), the $g$-processes play a negligible role, and the overall relative contribution of intervalley scattering is comparable to that in h-diamond, which is approximately 10\%.
Hence, intervalley scattering only plays a minor role in both phases around room temperature.

Therefore, the mobility difference between the two phases cannot be explained by either the interband scattering that involves the sh band or the intervalley scattering of electrons.
According to the results above, we will henceforth focus on intravalley scattering and restrict our analysis to the hh and lh bands.

Furthermore, we calculated carrier scattering induced by ionised impurities.
Here, the ionised impurities are modelled as randomly distributed Coulomb scattering centres~\cite{Leveillee2023}, without considering the detailed defect structures and their effect on the band structure or phonon dispersion.
According to Eq.~(S9), under a certain doping condition and temperature, the impurity-limited carrier mobility is related to the band curvature near the band edge and the screening.
Given the similar dielectric constants of both phases (see Table~S2), this mobility can serve as a measurement for the effect of band structure on carrier transport.
The charge state and concentration of ionised impurities were chosen to be 1 and $10^{17}$~cm$^{-3}$, respectively, and the impurity-limited carrier mobilities (without phonon scattering) at 300~K exhibit only small differences between h-diamond and c-diamond, as listed in Table~\ref{tab:ii-mob}. 
This result further confirms that the band structure plays similar roles in the two phases, suggesting that the much higher mobilities in h-diamond should be attributed to its weaker e-ph interaction.
It should be noted that this impurity-limited mobility here serves primarily as a conceptual probe to isolate band effects. 
In practical device architectures, such ionised impurity scattering can be effectively bypassed via spatial separation techniques, such as surface transfer doping~\cite{Jurgen2006} or modulation doping~\cite{Wang2022}, thereby preserving the exceptional phonon-limited mobilities of h-diamond.

\subsection{Origin of the higher hole mobility in h-diamond}

\begin{table}
\caption{\label{tab:mob-mode-h}Calculated average hole relaxation times $\langle\tau\rangle$ (in ps) for different phonon modes at 300~K, defined as $\langle\tau\rangle = \frac13\sum_\alpha m_\alpha\mu_\alpha / e$. Here, $\alpha$ represents the Cartesian direction, $m_\alpha$ the mobility effective mass, $\mu_\alpha$ the SERTA hole mobility limited by selected phonon modes, and $e$ the elementary charge. Corresponding ratios between h-diamond and c-diamond are provided in the final row.} 
\begin{ruledtabular}
    \begin{tabular}{ccccc}
                  & All modes & TA & LA & Optical \\
         \hline
        h-diamond & 18.3 & 65.9 & 76.7 & 55.0 \\
        c-diamond & 6.98 & 11.9 & 33.4 & 53.8 \\
        Ratio     & 2.62 & 5.55 & 2.29 & 1.02 \\
    \end{tabular}
\end{ruledtabular}
\end{table}

\begin{figure*}
    \centering
    \includegraphics[width=\linewidth]{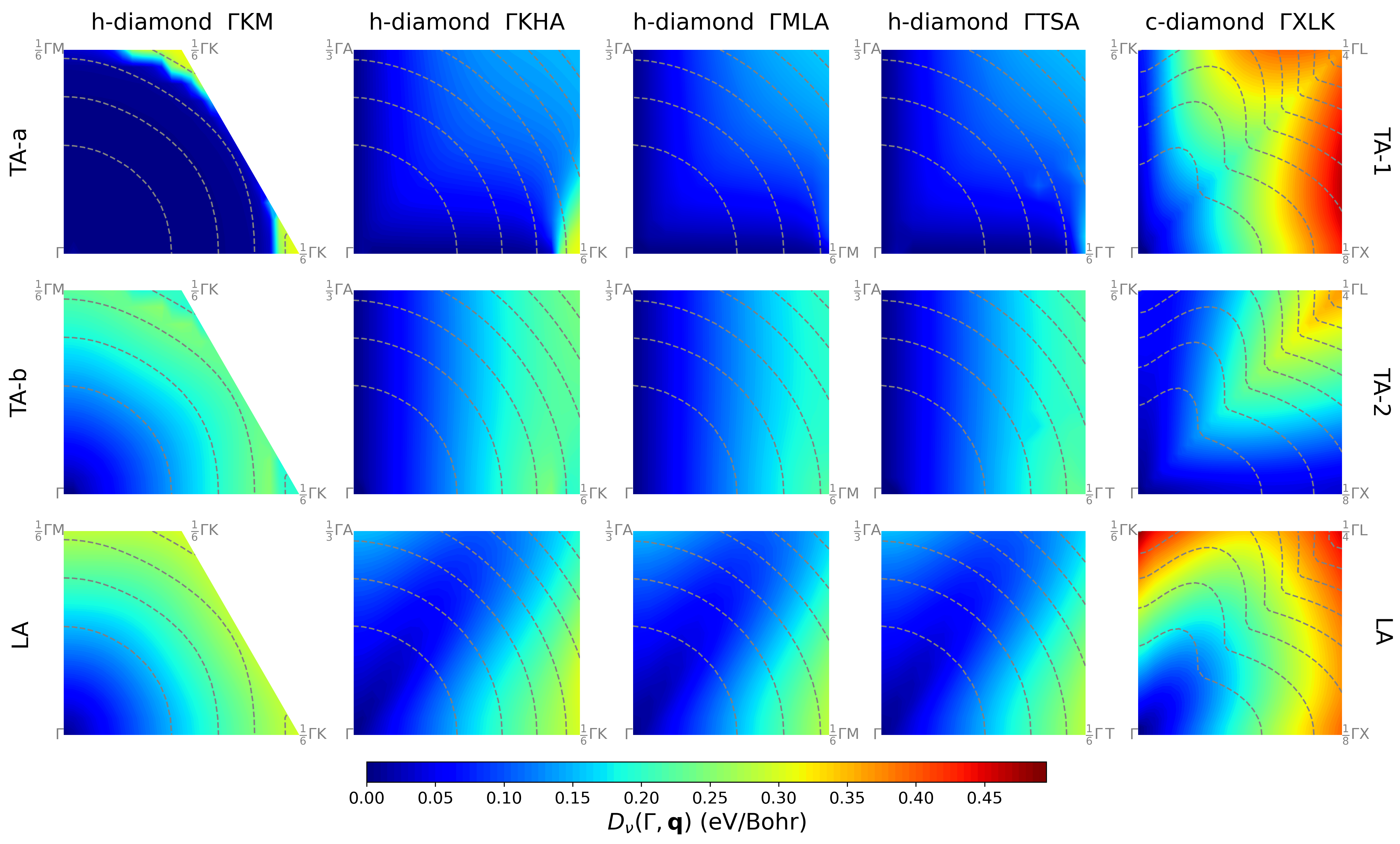}
    \caption{Distributions of the hole deformation potentials $D_\nu (\Gamma, \mathbf q)$ [averaged over the hh and lh bands, see Eq.~(S10)] for the three acoustic modes on selected planes in the BZ. For h-diamond, the $\Gamma KM$ plane ($k_z=0$) is perpendicular to the other three vertical planes. Alongside the high-symmetry $\Gamma KHA$ and $\Gamma MLA$ planes, the lower-symmetry $\Gamma TSA$ plane is included here to represent a more general case, where $T$ and $S$ denote the midpoints of the $KM$ and $HL$ paths, respectively. For c-diamond, a single plane perpendicular to the [$10\overline1$] direction is shown to incorporate the [010] $\Gamma X$, [101] $\Gamma K$, and [111] $\Gamma L$ directions. Grey dashed contours indicate the hole energy distribution of the hh band with an interval of 0.1~eV. Only a small region near the VBM is displayed, as high-energy states contribute negligibly to carrier transport.}
    \label{fig:dp-h-plane}
\end{figure*}

As shown in Fig.~\ref{fig:mob-freq}, ADP scattering is the predominant e-ph scattering mechanism near room temperature.
To quantify the contributions of individual phonon modes, we calculated the average hole relaxation time limited by specific modes at 300~K based on the self-energy relaxation time approximation (SERTA) mobility and the mobility effective mass.
To enable a straightforward comparison, we also calculated the ratio of the relaxation time in h-diamond to that in c-diamond for each phonon mode or mode combination.
A higher ratio indicates weaker hole scattering induced by corresponding modes in the hexagonal phase compared with the cubic phase, and vice versa.
As listed in Table~\ref{tab:mob-mode-h}, the transverse acoustic (TA)-limited hole relaxation time is shorter than the longitudinal acoustic (LA)- or optical-limited values in both phases.
This indicates that ADP scattering mediated by TA phonons dominates hole scattering around room temperature, in agreement with previous theoretical study on c-diamond~\cite{Ponce2021}.
Notably, the TA-limited hole relaxation time is significantly longer in h-diamond than in c-diamond, with a ratio that considerably exceeds the ratio of the total hole relaxation times.
The LA-limited hole relaxation time exhibits a similar trend, though with a lower ratio between the hexagonal and cubic phases.
In contrast, the optical-phonon-limited hole relaxation time in h-diamond is comparable to its counterpart in c-diamond.
Therefore, the higher hole mobility in h-diamond is primarily attributed to significantly weaker scattering induced by acoustic phonons, particularly the TA modes.
In the following, we comprehensively compare the behaviour of different phonon modes in the two phases, and elucidate the underlying mechanisms.

\begin{figure}
    \centering
    \includegraphics[width=\linewidth]{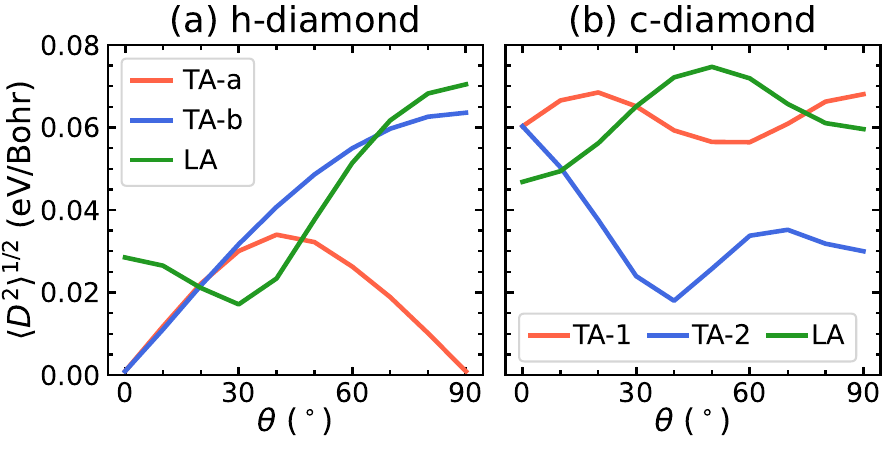}
    \caption{Root-mean-square acoustic-induced hole deformation potential $\langle D^2\rangle ^{1/2}$ (averaged over the azimuthal angle $\varphi$) as a function of the polar angle $\theta$. The phonon wavevector length is fixed at $0.005\times2\pi$~Bohr$^{-1}$. For (a) h-diamond, the $c$-axis serves as the $z$-axis of the polar coordinate system, where $\theta = 0$ and $90^\circ$ represent $\mathbf q\parallel c$ and $\mathbf q\perp c$, respectively. For (b) c-diamond, the [111] stacking direction is chosen as the $z$-axis, with $\theta$ denoting the angle between $\mathbf q$ and [111]. Due to lattice symmetry, only the $0 \leq \theta \leq 90^\circ$ range is displayed.}
    \label{fig:dp-h-angle}
\end{figure}

\begin{figure*}
    \centering
    \includegraphics[width=\linewidth]{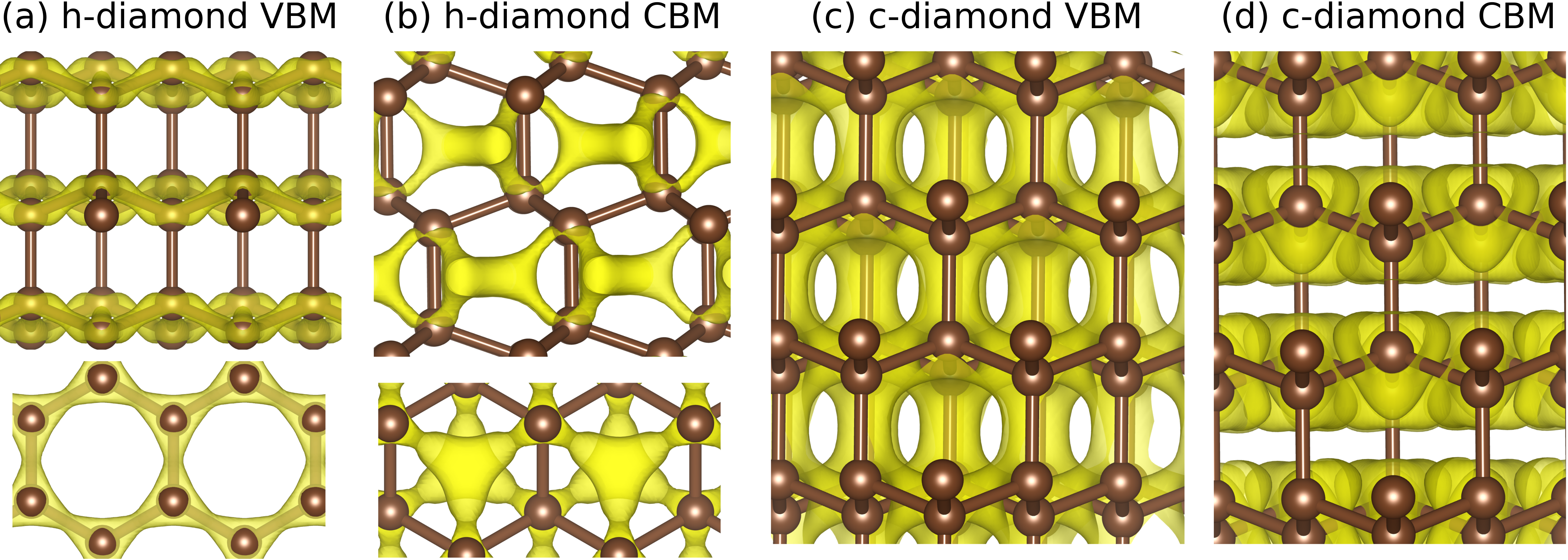}
    \caption{Real-space distributions of charge density at the VBM and CBM. The isosurfaces denote values of (a)~$3.5\times10^{-4}$, (b)~$1.8\times10^{-3}$, (c)~$7\times10^{-4}$, and (d)~$1.08\times10^{-3}$~a.u., respectively. Visualised using \textsc{vesta}~\cite{vesta}.}
    \label{fig:charge}
\end{figure*}

We first focus on the TA phonons in h-diamond and demonstrate that the e-ph selection rules play a crucial role.
To avoid any ambiguity, the two TA modes are explicitly referred to as TA-a and TA-b.
As shown in Fig.~\ref{fig:hex-TA}, the polarisation vector of TA-a lies within the plane defined by the $c$-axis and the wavevector $\mathbf q$ (for $\mathbf q \nparallel c$), whereas that of TA-b always lies within the $k_z=0$ plane.
For in-plane scattering between electronic states within the $k_z=0$ plane, because both the initial and final states near the VBM are odd under the $k_z=0$ mirror symmetry, phonons with odd parity with respect to this plane cannot induce such transitions.
Among the three acoustic modes, the TA-a mode, which manifests as out-of-plane vibrations [see Fig.~\ref{fig:hex-TA}(c)], is thus forbidden by this selection rule.
The selection rule at the VBM for scattering along the $c$-axis ($\mathbf q \parallel c$) is given by (the notation for irreducible representations follows that of Ref.~\onlinecite{Dresselhaus2008}; see also Fig.~S4)
\begin{equation}\label{eq:h-vb-outplane}
    \Gamma_5^+ \otimes \Delta_5 = \Delta_1 \oplus \Delta_2 \oplus \Delta_6.
\end{equation}
This indicates that both TA modes (with $\Delta_5$ symmetry) are forbidden.
This selection rule can be further extended to other scattering processes along the $\Gamma$-$A$ path, as the initial and final states both share the same $C_6$ rotational axis.

Nevertheless, scattering processes away from high-symmetry directions, including first-order transitions, can still have a significant impact on carrier mobility, as seen in systems like silicon~\cite{Ferry1976,Wang2011}.
To visualise the spatial impact of these selection rules on the scattering channels across the BZ, we calculated the distribution of the hole deformation potential~\cite{Ponce2021} [Eq.~(S10)] on several planes (Fig.~\ref{fig:dp-h-plane}).
For the TA-a mode, the hole deformation potential is zero across a wide region near the VBM on the $k_z = 0$ plane and along the $c$-axis, in line with the selection rules analysed above.
Moreover, in planes parallel to the $c$-axis, including the high-symmetry $\Gamma KHA$ and $\Gamma MLA$ planes, and the low-symmetry $\Gamma TSA$ plane, the hole deformation potential induced by TA-a remains consistently small.
This is attributed to the fact that e-ph matrix elements are symmetry-restricted to zero along the boundaries ($\mathbf{q} \parallel c$ and $\mathbf{q} \perp c$) of these planes.
This effect extends into the intermediate regions owing to the continuous and smooth spatial variation of the matrix elements.
To provide further quantification, we first express the phonon wavevector $\mathbf q$ in spherical polar coordinates
\begin{equation}
    \mathbf{q} (q,\theta, \varphi) = (q\sin\theta\cos\varphi, q\sin\theta\sin\varphi, q\cos\theta),
\end{equation}
and calculate the root-mean-square hole deformation potential averaged over the azimuthal angle $\varphi$ for a given wavevector length $q$, yielding $\left\langle D^2\right\rangle^{1/2}$ as a function of the polar angle $\theta$.
Fig.~\ref{fig:dp-h-angle} shows the relationship between $\left\langle D^2\right\rangle^{1/2}$ and $\theta$ at $q = 0.005\times2\pi$~Bohr$^{-1}$.
Similar trends are observed for larger wavevector lengths ($q = 0.010$ and $0.015\times2\pi$~Bohr$^{-1}$), with $\left\langle D^2\right\rangle^{1/2}$ scaling accordingly, which confirms that this behaviour holds for phonons across a wide range of wavevector lengths.
Although the hole deformation potential induced by the TA-a mode becomes non-zero when $0<\theta<90^\circ$, the restrictions at the boundaries ($\theta = 0$ and $\theta = 90^\circ$) suppress the peak value.
This demonstrates that the selection rules along the rotation axis and on the mirror plane exert a profound effect throughout a wide region near the VBM, substantially suppressing the scattering mediated by TA-a phonons.

For the TA-b mode, scattering within the $k_z=0$ plane is symmetry-allowed, and thus the hole deformation potential increases considerably with $\vert\mathbf q\vert$ (Fig.~\ref{fig:dp-h-plane}).
Although scattering along the $c$-axis is forbidden, the scattering channel opens as $\mathbf q$ deviates from it, leading to a monotonic increase in the total deformation potential as $\theta$ varies from 0 to 90$^\circ$ [see Fig.~\ref{fig:dp-h-angle}(a)].
Consequently, the TA-b mode contributes more significantly to hole scattering than the TA-a mode does.

\begin{figure}
    \centering
    \includegraphics[width=0.9\linewidth]{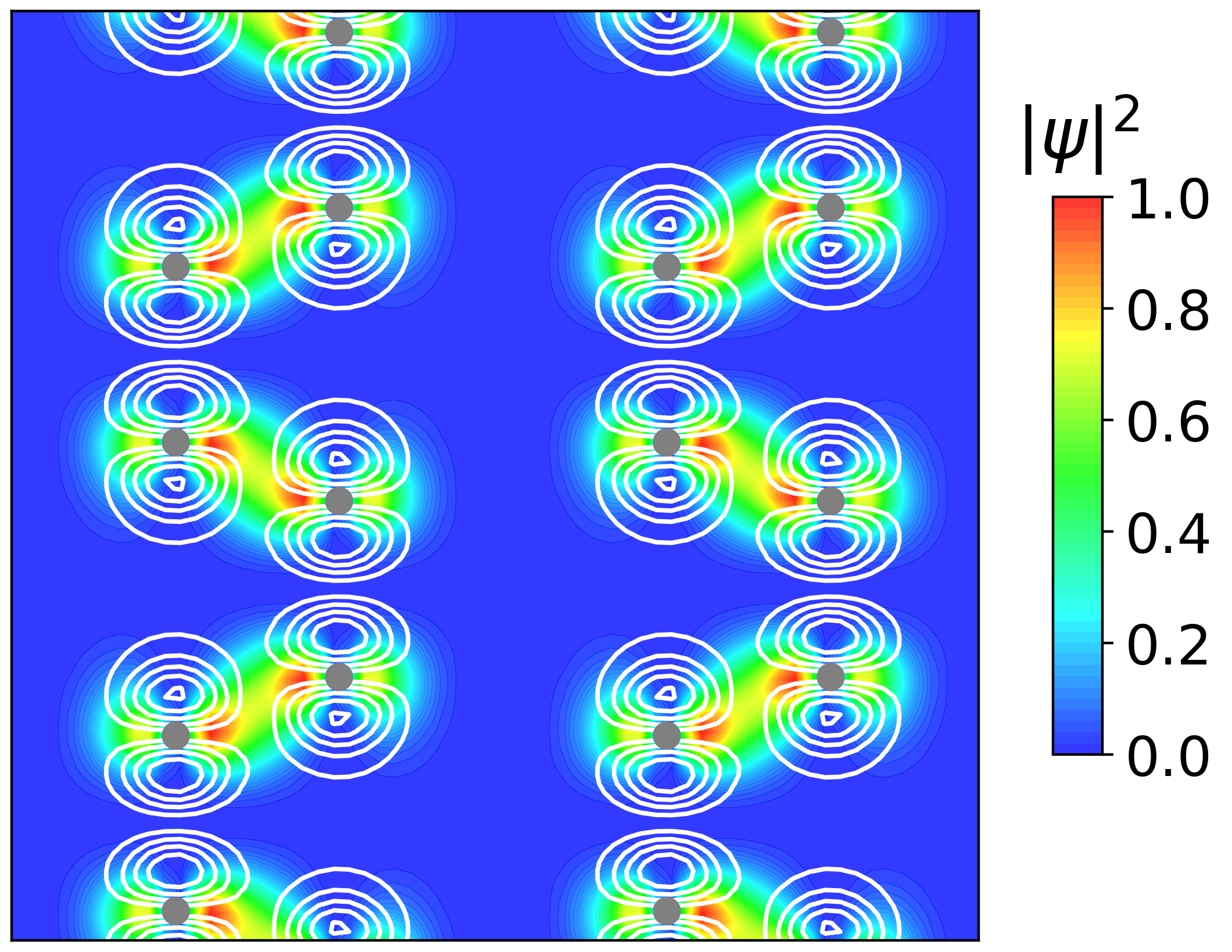}
    \caption{Distributions of the partial electronic density at the VBM (normalised, colour-filled) and the scattering potential induced by the LA phonon at $\mathbf q = 0.2~\Gamma\text{-}A$ (white contours) within the $(11\overline20)$ plane in h-diamond. Grey dots indicate the positions of C atoms. The spatial mismatch between the out-of-plane atomic vibration and the in-plane charge distribution leads to the suppression of scattering. }
    \label{fig:decoupling-h}
\end{figure}

Regarding LA phonons in h-diamond, hole scattering is symmetry-allowed in all directions.
Nevertheless, Figs.~\ref{fig:dp-h-plane} and \ref{fig:dp-h-angle}(a) indicate that the hole deformation potential depends strongly on $\theta$, peaking at $\theta=90^\circ$.
This trend arises from the in-plane charge density distribution of the VBM, which possesses distinct $p_x$-$p_y$ character [Fig.~\ref{fig:charge}(a)].
When $\theta \sim 0$, the LA mode primarily involves the stretching of C-C bonds perpendicular to the basal planes, resulting in relatively weak coupling with the in-plane distributed electronic wavefunction near the VBM.
This mechanism is visualised in Fig.~\ref{fig:decoupling-h}, which reveals an appreciable spatial mismatch between the partial electronic density (colour-filled) and the LA-induced scattering potentials (contours).

Whilst the hole deformation potentials shown in Figs.~\ref{fig:dp-h-plane} and \ref{fig:dp-h-angle} are calculated for initial states at the $\Gamma$ point, the impact of the selection rules extends well beyond the exact VBM point.
For example, the selection rules enforced by the $C_6$ axis and the $\sigma_h$ mirror plane remain strictly valid for scattering between electronic states located along this axis or within this plane, respectively.
Furthermore, we calculated the angular dependence of the deformation potential with the initial state shifted off the $\Gamma$ point, including two $\mathbf k$ points along the high-symmetry $\Gamma\text{-}A$ and $\Gamma\text{-}K$ paths, and one low-symmetry $\mathbf k$ point belonging to the $C_1$ little group.
As shown in Fig.~S5, although some  selection rules no longer strictly hold for these off-$\Gamma$ points, the angular dependence of all three $\mathbf k$ points remains almost identical to that at the $\Gamma$ point.
Consequently, the residual influence of these selection rules extends over a considerable region around the VBM.

The behaviour of acoustic phonons in h-diamond also rationalises the observed anisotropy and the difference between the Boltzmann transport equation (BTE) and SERTA results.
At 300~K, the mobility effective mass parallel to the $c$-axis is significantly larger than that perpendicular to it (Table~\ref{tab:mass}), leading to a lower SERTA mobility along the $c$-direction.
However, the BTE results indicate that the hole mobility parallel to the $c$-axis is slightly higher than that in the perpendicular directions.
This discrepancy arises from the fundamental assumption of the SERTA, namely, a complete loss of carrier momentum after each scattering event.
According to Fig.~\ref{fig:dp-h-angle}(a), scattering mediated by acoustic phonons is much weaker for $\mathbf q \parallel c$ than for $\mathbf q \perp c$.
Hence, forward scattering along the $c$-axis predominates, leading to high retention of the $c$-component of carrier velocity after a scattering event.
Consequently, the carrier mobility along the $c$-axis is enhanced despite its larger effective mass, resulting in a considerable underestimation of the SERTA mobility parallel to the $c$-axis [Fig.~S6(a)].

We now consider the behaviour of acoustic phonons in c-diamond.
In case the two TA branches are not degenerate, the low-energy and high-energy modes are labelled as TA-1 and TA-2, respectively.
The selection rules for e-ph scattering at the VBM point along high-symmetry paths are
\begin{align}\label{eq:c-vb-100}
    \langle100\rangle:\quad& \nonumber \\
    \Gamma_{25}^\prime \otimes \Delta_2^\prime & = \Delta_1 \oplus \Delta_5 = \mathrm{1TA+LA+1TO}, \nonumber \\
    \Gamma_{25}^\prime \otimes \Delta_5 &= \Delta_1 \oplus \Delta_1^\prime \oplus \Delta_2 \oplus \Delta_2^\prime \oplus \Delta_5 \nonumber \\ 
    & = \mathrm{1TA+LA+1TO+LO},
\end{align}

\begin{align}\label{eq:c-vb-110}
    \langle110\rangle: \quad \nonumber \\ \Gamma_{25}^\prime \otimes \Sigma_2 & = \Sigma_1 \oplus \Sigma_2 \oplus \Sigma_4 = \mathrm{TA\text{-}1 + LA + 2TO}, \nonumber \\
    \Gamma_{25}^\prime \otimes \Sigma_1 & = \Sigma_1 \oplus \Sigma_2 \oplus \Sigma_3 = \mathrm{TA\text{-}2 + LA + 2TO + LO},
\end{align}

\begin{align}\label{eq:c-vb-111}
    &\langle111\rangle: \nonumber\\ 
    &\Gamma_{25}^\prime \otimes \Lambda_3 = \Lambda_1 \oplus \Lambda_2 \oplus 2\Lambda_3 = \mathrm{2TA + LA + 2TO + LO},
\end{align}

\noindent where TO and LO represent transverse optical and longitudinal optical phonons, respectively.
Consequently, symmetry imposes fewer restrictions on acoustic modes in c-diamond.
Fig.~\ref{fig:dp-h-plane} also confirms that the hole deformation potential remains non-zero across almost all directions.
To enable a quantitative comparison between the cubic and hexagonal phases, we calculated the angular dependence of hole deformation potential [see Fig.~\ref{fig:dp-h-angle}(b)] using the [111] direction as the polar axis, consistent with the ABC stacking orientation.
In c-diamond, the overall effect of the TA-2 mode is comparable to that of the TA-b mode in h-diamond.
However, scattering induced by TA-1 phonons in c-diamond is substantially stronger than that induced by TA-a phonons in h-diamond.
Regarding the hole deformation potential of LA phonons, although its peak value is similar in both phases, it is smaller in the hexagonal phase at small $\sin\theta$ values, indicating a weaker overall contribution in h-diamond.

\begin{figure}
    \centering
    \includegraphics[width=\linewidth]{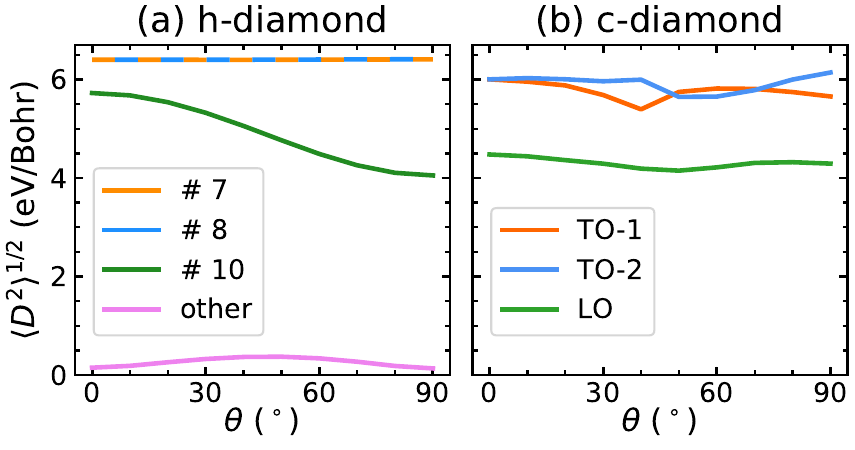}
    \caption{Root-mean-square optical-induced hole deformation potential $\langle D^2\rangle ^{1/2}$ as a function of $\theta$. The choice of the polar coordinate system and the phonon wavevector length are identical to those in Fig.~\ref{fig:dp-h-angle}. For (a) h-diamond, prominent optical modes are labelled by their energy indices, while ``other'' denotes the cumulative contribution $(\sum_\nu D_\nu^2)^{1/2}$ summed over all remaining optical modes except modes 7, 8, and 10. For (b) c-diamond, modes are classified into TO-1, TO-2, and LO according to their energies.}
    \label{fig:dp-h-angle-op}
\end{figure}

Beyond acoustic phonons, optical phonons also play a non-negligible role in carrier scattering.
In h-diamond, which contains four atoms in its primitive cell, there are more optical modes than in c-diamond, which could potentially provide additional scattering channels.
Nevertheless, not all optical modes make a substantial contribution to hole scattering in h-diamond.
To provide a quantitative description, we calculated the angular dependence of the hole deformation potentials for all optical modes.
Given the complexity of the optical vibration patterns in h-diamond, the modes are labelled according to their ascending energy order near the BZ centre, with the lowest and highest optical branches denoted as 4 and 12, respectively.
As shown in Fig.~\ref{fig:dp-h-angle-op}(a), only three optical modes in h-diamond exhibit non-negligible hole deformation potentials.
The discrepancy in the total numbers of optical modes and active optical scattering channels in h-diamond can be understood according to selection rules.
According to Table~S3, modes 7, 8, and 10 are the only three optical branches free from symmetry restrictions.
In addition to selection rules, the spatial distribution patterns of electronic wavefunctions and phonon-induced scattering potentials also play an appreciable role.
Compared with modes 9, 11, and 12, which are forbidden for hole scattering at both $\theta=0$ and $90^\circ$, modes 4, 5, and 6 are restricted at only one boundary, resulting in a weaker suppression effect.
However, these modes behave as interlayer shear (modes 4 and 5) or breathing (mode 6) vibrations, mainly involving the shear or stretching deformation of out-of-plane bonds.
This leads to weak coupling with the in-plane distributed charge density near the $\Gamma$ point, thereby suppressing e-ph scattering. 

Since both the number of effective optical scattering channels and the corresponding hole deformation potentials are comparable in the two phases, the average optical-limited hole relaxation time in h-diamond is almost identical to that in c-diamond (Table~\ref{tab:mob-mode-h}).
On the other hand, their high energies result in small Bose-Einstein occupation numbers near room temperature, and thus optical phonons only play a minor role compared with acoustic modes.
Nonetheless, at elevated temperatures, the role of optical modes becomes more appreciable.
Therefore, the relative difference in hole mobility between the two phases diminishes at high temperatures (Fig.~\ref{fig:mob-temp}).
This temperature-dependent activation also significantly alters the hole mobility anisotropy.
As analysed previously, the acoustic scattering is significantly anisotropic: scattering is much weaker near $\theta = 0$ than at $\theta = 90^\circ$ [see Fig.~\ref{fig:dp-h-angle}(a)], which enhances the mobility along the $c$-direction.
In contrast, as shown in Fig.~\ref{fig:dp-h-angle-op}(a), the optical-induced deformation potentials are less sensitive to $\theta$, and are even slightly larger at $\theta = 0$ than at $\theta = 90^\circ$.
Consequently, as optical modes become thermally activated and begin to outweigh acoustic phonons, the hole mobility perpendicular to the $c$-axis surpasses the $c$-component, consistent with the effective mass anisotropy.
The discrepancy between the BTE and SERTA results also decreases.
From 300 to 700~K, the underestimation of the hole mobility along the $c$-axis drops from 16\% to only 3.5\% [Fig.~S6(a)].

In brief, the high room-temperature hole mobility in h-diamond primarily arises from symmetry-induced selection rules that substantially suppress acoustic scattering channels, particularly the TA-a mode.
The in-plane charge distribution at the VBM also plays an appreciable role by weakening the coupling with both out-of-plane LA vibrations and the interlayer optical modes.

\subsection{Origin of the higher electron mobility in h-diamond}

We now turn our attention to the conduction band, following the same framework as established in the previous section.

The room-temperature electron relaxation times limited by specific phonon modes are listed in Table~\ref{tab:mob-mode-e}.
TA-induced scattering predominates in c-diamond, as indicated by its shortest relaxation time among all modes.
In contrast, TA scattering contributes negligibly in h-diamond, exhibiting a relaxation time two orders of magnitude longer than that in c-diamond.
The electron relaxation times limited by LA or optical modes are also longer in the hexagonal phase than those in the cubic phase.
The exceptionally high electron mobility in h-diamond thus primarily arises from the suppression of TA scattering, with weaker scattering mediated by LA and optical modes further contributing to this enhancement.

\begin{figure*}
    \centering
    \includegraphics[width=\linewidth]{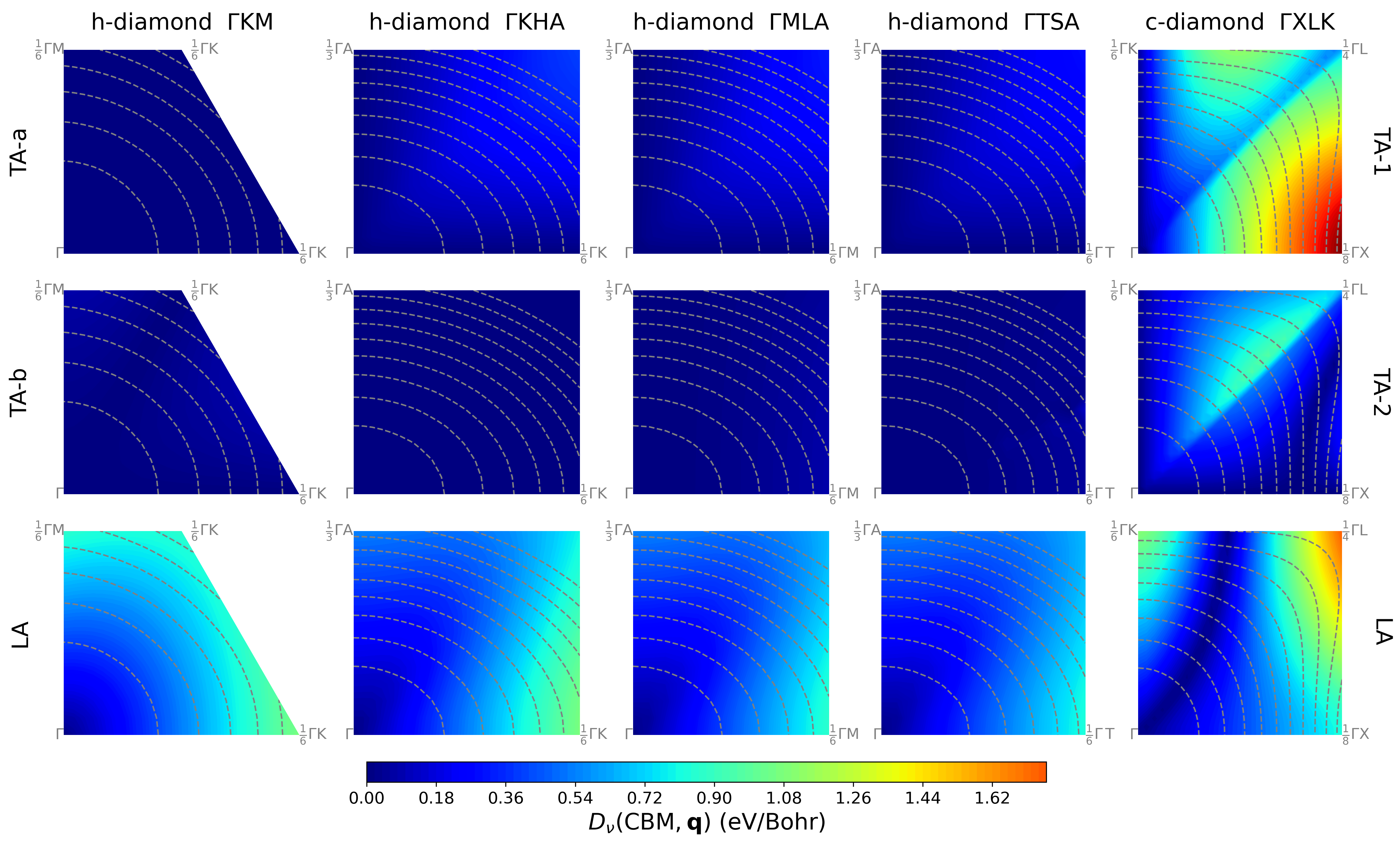}
    \caption{Distributions of the electron deformation potentials $D_\nu (\mathrm{CBM}, \mathbf q)$ for the three acoustic modes on selected planes in the BZ. The high-symmetry path points indicate the position of phonon wavevectors. The planes and displayed regions shown here are the same as those in Fig.~\ref{fig:dp-h-plane}. Grey dashed contours indicate the distribution of electronic energies with an interval of 0.1~eV.}
    \label{fig:dp-e-plane}
\end{figure*}

\begin{figure}
    \centering
    \includegraphics[width=\linewidth]{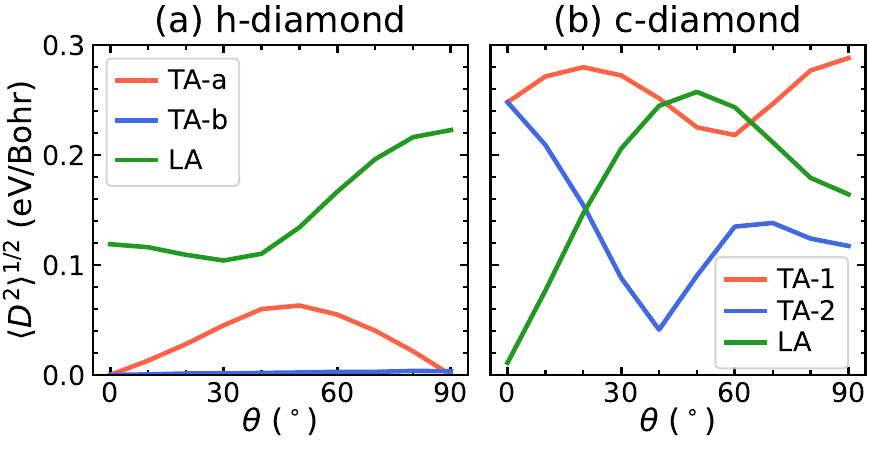}
    \caption{Root-mean-square acoustic-induced electron deformation potential $\langle D^2\rangle ^{1/2}$ as a function of $\theta$. The choice of the polar coordinate system and the phonon wavevector length are identical to those in Fig.~\ref{fig:dp-h-angle}.}
    \label{fig:dp-e-angle}
\end{figure}

\begin{table} 
\caption{\label{tab:mob-mode-e}Calculated average electron relaxation times $\langle\tau\rangle$ (in ps) for different phonon modes at 300~K, defined as $\langle\tau\rangle = \frac13\sum_\alpha m_\alpha\mu_\alpha / e$. Here, $\alpha$ represents the Cartesian direction, $m_\alpha$ the mobility effective mass, $\mu_\alpha$ the SERTA electron mobility limited by selected phonon modes, and $e$ the elementary charge. Corresponding ratios between h-diamond and c-diamond are provided in the final row.} 
\begin{ruledtabular}
    \begin{tabular}{ccccc}
                  & All modes & TA & LA & Optical \\
         \hline
        h-diamond & 34.9 & 727 & 49.2 & 172 \\
        c-diamond & 4.40 & 6.60 & 18.9 & 62.1 \\
        Ratio     & 7.94 & 110 & 2.61 & 2.77 \\
    \end{tabular}
\end{ruledtabular}
\end{table}

\begin{figure*}
    \centering
    \includegraphics[width=\linewidth]{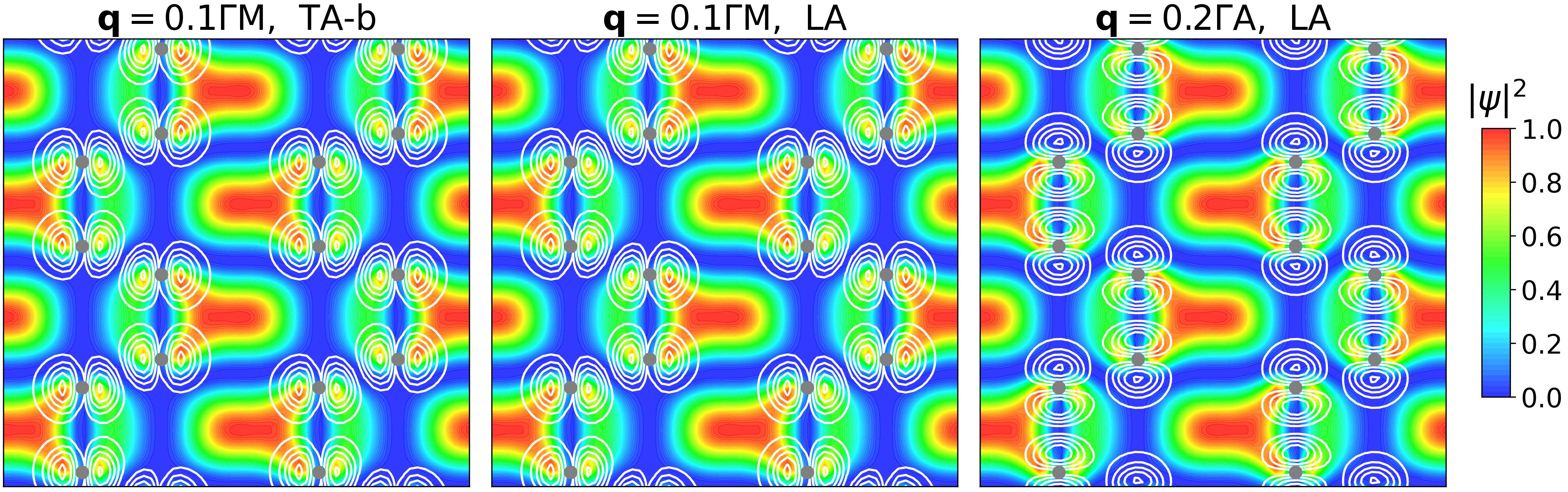}
    \caption{Distributions of the partial electronic density at the CBM (normalised, colour-filled) and the scattering potential induced by the different phonons within the $(11\overline20)$ plane in h-diamond. Grey dots indicate the positions of C atoms.}
    \label{fig:decoupling-e}
\end{figure*}

We first focus on TA-induced scattering in h-diamond under selection rules.
The little group of the CBM (at the $K$ points) is $D_{3h}$, which includes $C_3$ and $\sigma_h$ symmetry operations.
Hence, several selection rules for electrons are similar to those for holes.
Specifically, for scattering between electronic states along the $K$-$H$ path, the selection rule is given by
\begin{equation}\label{eq:hex-e-z}
    K_2^+ \otimes P_2 = \Delta_1 \oplus \Delta_4,
\end{equation}
which forbids both TA modes.
For scattering between electronic states within the $k_z=0$ plane, the TA-a mode is forbidden owing to its odd parity with respect to this mirror plane.
Consequently, the electron deformation potential induced by the TA-a mode exhibits a distribution pattern and angular dependence similar to those of the hole deformation potential, as shown in Figs.~\ref{fig:dp-e-plane} and \ref{fig:dp-e-angle}(a), thereby resulting in weak scattering.

Notably, the selection rules for the TA-b mode are substantially more restrictive for electrons than for holes.
For holes, scattering induced by the TA-b mode is symmetry-forbidden only along the $c$-axis, yielding an appreciable deformation potential when the phonon wavevector is nearly parallel to the $k_z = 0$ plane [see Figs.~\ref{fig:dp-h-plane} and \ref{fig:dp-h-angle}(a)].
For electrons, however, scattering between states within the same $\Gamma KHA$ plane is forbidden, as both the initial and final states have odd parity with respect to this $\sigma_v$ plane, and the TA-b mode itself also has odd parity.
Due to the hexagonal crystallographic symmetry, these vertical symmetry planes intersect along the $c$-axis at 60$^\circ$ intervals, effectively partitioning the three-dimensional $\mathbf q$-space into narrow regions.
Scattering mediated by TA-b phonons within these regions is thus significantly suppressed by the symmetry-imposed restrictions on the closely spaced boundaries.
Consequently, the electron deformation potential corresponding to the TA-b mode is nearly negligible across the entire phase space, as confirmed by Figs.~\ref{fig:dp-e-plane} and \ref{fig:dp-e-angle}(a).

We also calculated the angular dependence of the electron deformation potentials for initial states shifted off the CBM point, as shown in Fig.~S7. 
A similar trend is observed, indicating that the residual suppression effect of the selection rules remains effective across an appreciable region near the CBM.

Nevertheless, selection rules alone cannot fully account for the observed difference in acoustic scattering.
Whilst the TA-scattering in h-diamond is subject to selection rules, the same holds true for c-diamond.
For example, the CBM of c-diamond is located at the $\Delta$ points along the $\langle100\rangle$ directions, where two $\sigma_v$ and two $\sigma_d$ mirrors exist.
For scattering within the same mirror plane, the out-of-plane TA mode is forbidden.
However, as shown in Fig.~\ref{fig:dp-e-angle}(b), the electron deformation potentials are substantially non-zero for both TA modes in c-diamond.
Moreover, the LA mode lacks symmetry-enforced protection in either phase, yet the LA-limited average electron relaxation time is 2.61 times longer in h-diamond.
These observations can be rationalised by the electronic charge distribution.
As shown in Fig.~\ref{fig:charge}(b), the CBM state in h-diamond is composed of antibonding $p_x$-$p_y$ orbitals, in line with the $D_{3h}$ little-group symmetry at the $K$ points.
The corresponding charge density is primarily distributed within the ``cages'' formed by C atoms and C-C bonds.
Since h-diamond is non-polar and  thus lacks long-range Fr\"ohlich interactions, phonon-induced scattering potentials are typically concentrated near atoms and bonds, resulting in a spatial mismatch with the CBM wavefunction, as visualised in Fig.~\ref{fig:decoupling-e}.
Because the e-ph matrix elements are essentially the overlap integrals between electronic wavefunctions and scattering potentials, this partial ``decoupling'' considerably decreases their values, thereby reducing the scattering rate.
Hence, the negligible electron deformation potential induced by TA-b phonons [Fig.~\ref{fig:dp-e-angle}(a)] stems from a combination of selection rules in specific directions and the e-ph decoupling, with the latter considerably suppressing the first-order processes.
For LA phonons, the spatial mismatch between the charge density and the scattering potential at $\theta=0$ is more significant than that at $\theta = 90^\circ$ (see Fig.~\ref{fig:decoupling-e}), which is consistent with the angular dependence of the LA-induced deformation potential shown in Fig.~\ref{fig:dp-e-angle}(a).
In contrast, the CBM wavefunction in c-diamond consists of antibonding $sp^3$ orbitals concentrated near the atoms and covalent bonds [see Fig.~\ref{fig:charge}(d)], which couple strongly with scattering potentials (Fig.~S8).
Consequently, the electron deformation potentials are systematically lower in h-diamond for both TA and LA modes, as shown in Fig.~\ref{fig:dp-e-angle}.

\begin{figure}
    \centering
    \includegraphics[width=\linewidth]{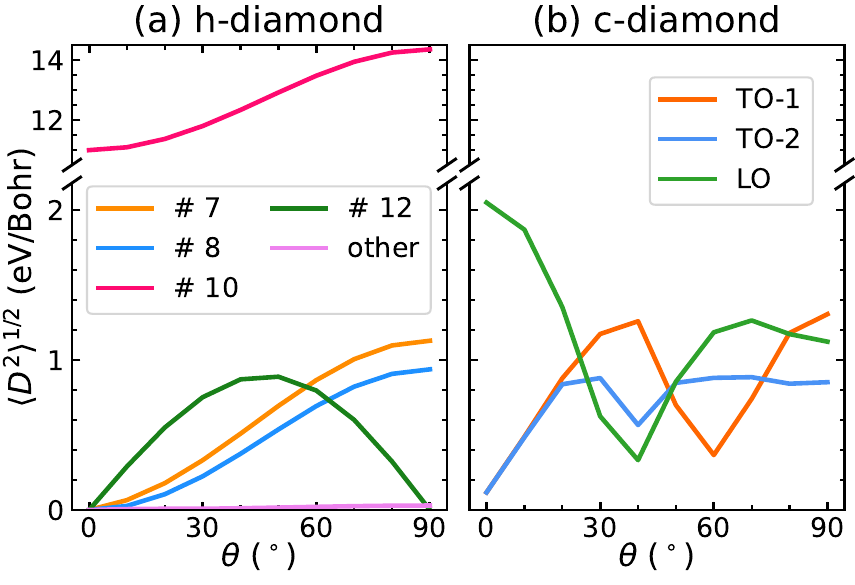}
    \caption{Root-mean-square optical-induced electron deformation potential $\langle D^2\rangle ^{1/2}$ as a function of $\theta$. The choice of the polar coordinate system and the phonon wavevector length are identical to those in Fig.~\ref{fig:dp-h-angle}. For (a) h-diamond, prominent optical modes are labelled by their energy indices, while ``other'' denotes the cumulative contribution $(\sum_\nu D_\nu^2)^{1/2}$ summed over all remaining optical modes except modes 7, 8, 10, and 12. For (b) c-diamond, modes are classified into TO-1, TO-2, and LO according to their energies.}
    \label{fig:dp-e-angle-op}
\end{figure}

For optical phonons, only a few modes in h-diamond make a non-negligible contribution.
As shown in Fig.~\ref{fig:dp-e-angle-op}(a), the deformation potentials induced by modes 7, 8, and 12, whose angular dependences match the selection rules summarised in Table~S3, make an overall contribution comparable to that in c-diamond.
Nonetheless, mode 10 exhibits substantially higher deformation potentials than either other optical branches in h-diamond, or those in c-diamond.
This can also be attributed to the different selection rules.
Mode 10 in h-diamond belongs to a fully symmetric representation, and is thereby not restricted by selection rules in any direction.
By comparison, scattering mediated by other optical modes in both phases is symmetry-forbidden in specific directions.
Additionally, mode 10 in h-diamond is characterised by relative vibrations of all neighbouring atoms that lead to a considerable distortion of the cages.
This structural deformation strongly perturbs the potential environment within these cages, thereby enhancing the coupling with the locally distributed charge.
The significant deformation potential induced by mode 10 appears to contradict the longer optical-limited electron relaxation time in h-diamond, which can be rationalised by examining the competition between intravalley and intervalley scattering.
As discussed previously, intervalley scattering contributes approximately 10\% to the room-temperature electron mobility reduction in both phases, and thus our analysis of selection rules and deformation potentials focuses primarily on intravalley processes.
However, for optical modes, the effect of intervalley scattering is more pronounced than that of intravalley scattering.
As shown in Fig.~\ref{fig:mob-freq}, the mobility reduction index $M(\hbar\omega)$ has two prominent peaks in the high-energy regime for both phases.
Whilst the higher-energy peak stems from optical-phonon-mediated intravalley scattering, the lower-energy peak does not correspond to any phonon branch near the BZ centre, indicating that it should arise from intervalley scattering induced by optical phonons with $\mathbf q$ near the BZ boundary.
At 300~K, the lower-energy peak outweighs the higher-energy one, demonstrating that intervalley scattering outweighs intravalley scattering for optical modes in both phases.
Consequently, the optical-limited electron relaxation times at room temperature are mainly determined by intervalley scattering.
Because the relative contributions of intervalley scattering to mobility reduction are comparable in the two phases, and given the much higher electron mobility in h-diamond, the absolute strength of intervalley scattering is weaker in the hexagonal phase, thereby leading to a longer optical-limited electron relaxation time.
At elevated temperatures, the contribution of optical-induced intravalley scattering is substantially enhanced in h-diamond, whereas it remains secondary to intervalley scattering in c-diamond (Fig.~\ref{fig:mob-freq}).
As a consequence, the relative difference in electron mobility between the two phases decreases with increasing temperature.
For example, at 300~K, the electron mobility in h-diamond obtained by the BTE is 5.7 and 13.2 times that in c-diamond along the $\perp c$ and $\parallel c$ directions, respectively, whereas at 700~K, these ratios drop to 3.8 and 7.6, respectively.

In brief, selection rules significantly suppress electron scattering induced by both TA modes in h-diamond.
The spatial mismatch between the charge density and the scattering potentials serves as another crucial mechanism that weakens electron scattering.
Although the fully symmetric optical mode induces a significant deformation potential, it contributes minimally near room temperature, likely due to its high phonon energy.

\section{Discussion}
In summary, we have demonstrated the exceptionally high carrier mobilities in h-diamond and elucidated their physical origins.
ADP scattering mediated by TA phonons, which is the predominant scattering mechanism in c-diamond around room temperature, is significantly suppressed in h-diamond by selection rules stemming from its unique symmetry.
For example, the $k_z=0$ mirror plane, which is absent in the common wurtzite structure despite its hexagonal lattice and AB stacking sequence, plays a crucial role in this suppression.
Additionally, the spatial distribution pattern of the carrier wavefunction serves as another crucial factor.
The in-plane distribution of the VBM wavefunction in h-diamond leads to weak interactions with both the LA phonons propagating along the out-of-plane direction, and the additional interlayer shear and breathing modes.
For electrons, the CBM wavefunction formed by antibonding $p_x$-$p_y$ orbitals is primarily distributed within the interstitial ``cages'' of the h-diamond lattice, resulting in a weak spatial overlap with phonon-induced scattering potentials, which markedly reduces the electron-phonon matrix elements.
Other factors, including the effective mass, are found to be comparable between the two phases, thus contributing negligibly to their mobility difference.
We also elucidate the temperature dependence of carrier mobility, which deviates from the power-law behaviour due to the increased occupation of high-energy optical phonons at elevated temperatures.
Our findings indicate that h-diamond is a highly promising candidate for high-frequency and high-speed applications.
Furthermore, by revealing the significance of symmetry-enforced selection rules and real-space electron-phonon ``decoupling'', this work offers fresh insights into the mechanisms governing high carrier mobility in semiconductors.

\section{Methods}

\subsection{DFT and DFPT calculations}

Geometry optimisation and phonon calculations were performed using the \textsc{quantum espresso} software~\cite{qe2009,qe2017,qe2020}.
The Broyden-Fletcher-Goldfarb-Shanno (BFGS) algorithm was utilised for geometry optimisation, with the convergence thresholds for energy, force, and stress set to $10^{-9}$~Ry, $10^{-6}$~Ry/Bohr, and 0.05~kbar, respectively.
The relaxed lattice parameters are listed in Table~S2.
Phonon calculations were based on density-functional perturbation theory (DFPT), with a convergence threshold of $10^{-17}$.
Optimised norm-conserving Vanderbilt pseudopotentials~\cite{oncvpsp} with full relativistic treatment from the \textsc{pseudo dojo} project~\cite{dojo} were used, and the plane-wave cutoff energy was chosen to be 100~Ry.
The PBEsol functional~\cite{pbesol} was applied for the approximation of exchange-correlation interaction.

\subsection{$GW$ band structure calculations}

In order to obtain accurate $\varepsilon$-$\mathbf k$ dispersion relationships, quasi-particle eigenvalues were calculated using the one-shot $G_0W_0$ method, as implemented in the \textsc{yambo} software~\cite{yambo2009,yambo2019}. 
The cutoff energies of the exchange self-energy and dielectric function were chosen to be 60 and 22~Ry, respectively.
The plasmon-pole model was employed with a plasma energy of 2~Ry.
The numbers of bands included in the calculations of the dielectric function and correlation self-energy scaled as $40n_\mathrm{a}n_\mathrm{s}$ and $60n_\mathrm{a}n_\mathrm{s}$, respectively, where $n_\mathrm a$ is the number of atoms in the unit cell, and $n_\mathrm{s}$ = 1 and 2 for the cases without and with spin-orbit coupling (SOC) activated, respectively.

The band structures and DOS were then obtained based on the maximally localised Wannier functions (MLWFs)~\cite{wannier1997,wannier2001} as implemented in the \textsc{wannier90} code~\cite{wannier90}.
The valence bands and conduction bands were treated separately during the Wannierisation process.
The initial projections were chosen as the $sp^3$ orbitals of half of the C atoms in the unit cell.
Specifically, one C atom was chosen for c-diamond, whereas for h-diamond, two C atoms that are not directly bonded to each other were selected.

\subsection{e-ph scattering and mobility calculations}
For the calculations of e-ph scattering and carrier mobility, the electronic Hamiltonians, dynamical matrices, and e-ph matrix elements were interpolated onto ultrafine $\mathbf k$ and $\mathbf q$ grids based on MLWFs using the \textsc{epw} code~\cite{epw2016,epw2023}.
The energy window was set to 0.3~eV (relative to the band edge), and the grid sizes are listed in Table~S4 of the SI, with the results of convergence tests provided in Figs.~S9 and S10.
To examine the quality of the Wannier interpolation, we calculated and compared the total deformation potentials~\cite{Ponce2021} with direct DFPT results, which exhibit excellent agreement (Fig.~S11).
The SOC was found to have a negligible effect on electron mobility and was thus only considered for valence bands.
To ensure the accurate description of long-range interactions, the effect of dynamical quadrupoles was included, which were calculated based on DFPT using the \textsc{abinit} software~\cite{abinit2020a,abinit2020b}.
We also calculated the dynamical quadrupole tensors using the implementation available in the \textsc{quantum espresso} package~\cite{Macheda2024}, and the results agree well with those obtained using \textsc{abinit}~(Table S2).

For cross-verification, carrier mobilities were also calculated using the \textsc{abinit} software~\cite{abinit2020a,abinit2020b}. 
The electronic states were obtained directly from non-self-consistent calculations, and Fourier interpolations were used for the scattering potentials.
The results are listed in Table~S1.

\subsection{MD simulation}

In order to examine the thermal stability of h-diamond, we performed an MD simulation in the $NPT$ ensemble at 1000~K and 1~atm.
The time step was chosen to be 1~fs, and the MD simulation ran for 10,000 steps (\textit{i.e.}, 10~ps).
Forces and pressures were calculated based on DFT using the Gaussian and plane wave (GPW) method as implemented in the \textsc{cp2k} code~\cite{cp2k}, employing the DZVP-MOLOPT basis set.
The cutoff energy and the relative cutoff energy were set to 300 and 50~Ry, respectively.
The exchange-correlation interaction was approximated by the PBEsol functional~\cite{pbesol}.
A $8\times8\times5$ supercell containing 1280 atoms was used, and the BZ was sampled with a single $\Gamma$-point.

\section*{Data availability}
The data files that support the findings of this work are available on the Materials Cloud Archive (https://doi.org/10.24435/materialscloud:rh-y0).

\section*{Code availability}
All computational codes used in this work (\textsc{quantum espresso}, \textsc{epw}, \textsc{abinit}, \textsc{yambo}, \textsc{wannier90}, and \textsc{cp2k}) are open-source and available online.

\section*{Competing interests}
The authors declare no competing interests.

\section*{Author contributions}
Z. H., S. G., and M. C. conceptualised and designed this study.
Z. H. performed the calculations, analysed the results, and wrote the initial draft.
S. G. and M. C. edited the manuscript and supervised the project.
All authors reviewed and approved the final version of the manuscript.

\begin{acknowledgments}
This work was supported by research funds from Shanghai Advanced Silicon Technology Co., Ltd., and also the Natural Science Foundation of Shanghai (Grant No. 23ZR1403300).
\end{acknowledgments}

%\nocite{*}
\bibliography{ref}

\end{document}